\documentclass[a4paper, 11pt, twoside]{amsart}

\usepackage{graphicx}
\usepackage{amsmath}
\usepackage{fancyhdr}

\title[The Markov-modulated sequentially Markov
coalescent]{Towards more realistic models of genomes in populations:
  the Markov-modulated sequentially Markov
  coalescent}

\author{Julien~Y.~Dutheil}

\email[J.~Y.~Dutheil]{dutheil@@evolbio.mpg.de}

\address{RG Molecular Systems Evolution, Max Planck Institute for Evolutionary Biology, August-Thienemann-Str. 2, 24306 Plön, GERMANY}

\begin{document}

\maketitle

The development of coalescent theory paved the way to statistical
inference from population genetic data. In the genomic era, however,
coalescent models are limited due to the complexity of the underlying
ancestral recombination graph. The sequentially Markov coalescent
(SMC) is a heuristic that enables the modelling of complete genomes
under the coalescent framework. While it empowers the inference of
detailed demographic history of a population from as few as one
diploid genome, current implementations of the SMC make unrealistic
assumptions about the homogeneity of the coalescent process along the
genome, ignoring the intrinsic spatial variability of parameters such
as the recombination rate. Here, I review the historical developments
of SMC models and discuss the evidence for parameter heterogeneity. I
then survey approaches to handle this heterogeneity, focusing on a
recently developed extension of the SMC.

\section{Modelling the evolution of genomes in populations}

When modelling the evolution of large genomic sequences at the
population level, in particular for sexually reproducing species, a
key biological mechanism to account for is meiotic recombination,
which shuffles genetic material at each generation. We first introduce
the concept of the ancestral recombination graph, needed to represent
the complete genealogy of a sample undergoing recombination. We then
review the statistical approaches used to fit models accounting for
recombination to population genomics data.

\subsection{The ancestral recombination graph}

\begin{figure}[p]
\begin{center}
  \includegraphics[width=0.9\textwidth]{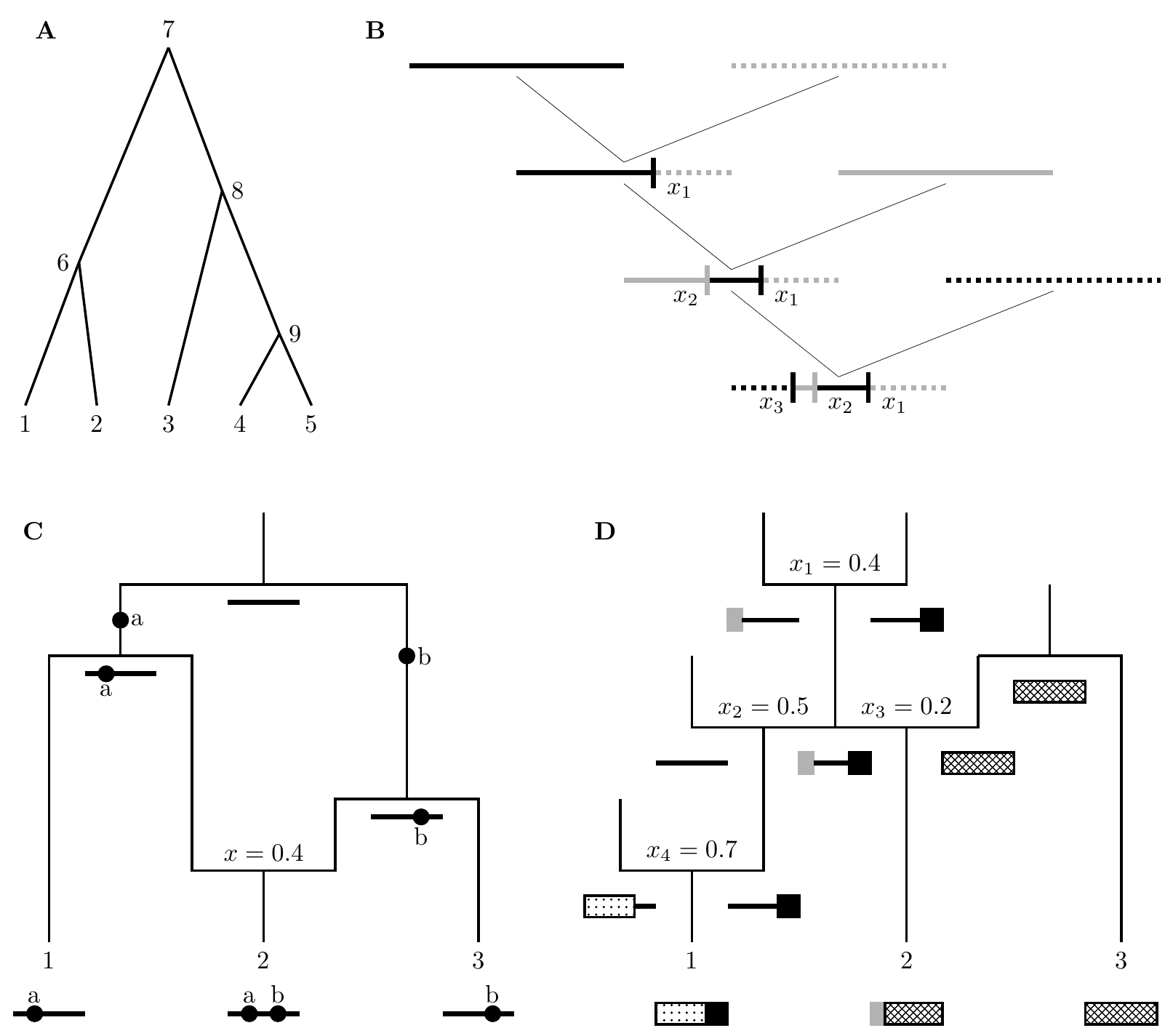}
\end{center}
\caption{\label{JD-fig1} Genealogies and recombination: relationships
  between individuals and along sequences.  A) example genealogy of a
  sample of five individuals at a given position in the sequences,
  under a Kingman coalescent. Tip nodes represent the samples (1-5)
  and inner nodes their common ancestors (6-9). B) illustration of the
  recombination tree resulting from a process without coalescence
  (large number approximation): two chromosomes (solid black and
  dotted grey) are paired during sexual reproduction and exchange
  segments at a breakpoint $x_1$. At the next generation, the
  descendant sequence recombines with another sequence (solid grey) at
  another breakpoint $x_2$, etc. The sequence of any sampled
  individual is therefore a mosaic of segments with distinct ancestors
  separated by a series of breakpoints ($x_1$, $x_2$, $x_3$). C) a
  simple ancestral recombination graph (ARG) representing the
  genealogy of a sample of size three with one recombination
  event. The ARG is a combination of a coalescence tree (as in A) and
  a recombination tree (as in B). For ease of interpretation, two
  mutation events, $\bullet a$ and $\bullet b$ have been added. The
  relative coordinate of the recombination event is also indicated:
  $x = 0.4$, assuming a total sequence length of 1. D) partial graph
  showing the different classes of recombination events. Ancestral
  segments are depicted as filled rectangles, while non-ancestral
  segments are shown as simple lines (subfigure created after Figure 1
  in \cite{JD-Marjoram2006}).  }
\end{figure}

The evolution of the set of sequences carried by all individuals
forming a population, generation after generation, can be modelled by
a stochastic process, where each individual leaves a variable number
of descendants in the next generation. As a result, at any position of
the sequence, the \emph{genealogy}\index{genealogy} of a sample of $n$
individuals can be described by a tree (Figure~\ref{JD-fig1}A)
\cite{JD-Kingman1982}. The tips of the tree represent the sampled
individuals and the inner nodes their common ancestors. In the case of
sexually reproducing organisms, which will be the focus of this
chapter, the genealogy is not identical for every position in the
sequence. During \emph{sexual reproduction}\index{sexual reproduction},
two individuals contribute part of their sequence to
their descendant(s) in the next generation. The mechanism of
\emph{recombination}\index{recombination} is responsible for randomly
sampling the new sequence from the two parental ones (Figure
\ref{JD-fig1}B). How often and where the recombination points occur
will be discussed in Section~\ref{JD-section-het}. The consequences of
the recombination process can be stated as: (1) the genealogy of the
sequence on the left of a recombination breakpoint potentially differs
from the genealogy of the sequence on the right, (2) the genealogy at
two positions in the sequence are more likely to differ as the
distance between the two points is large, and (3) the genealogy of the
complete sequence can no longer be described by a single tree, but by
a collection of such trees and associated breakpoints. This tree and
breakpoints collection can be represented as a single graph, called
the \emph{ancestral recombination graph (ARG)}\index{ancestral!recombination graph} (Figure \ref{JD-fig1}C)
\cite{JD-Griffiths1996}. The complexity of the ARG grows with the
number of individuals (which dictates the size of the underlying
trees) and the number of recombination events (which determines how
many trees are needed to represent the history of the full
sequences). The ARG represents the complete history of the sampled
individuals, where the trees at each position (referred to as the
``marginal genealogies'')are embedded \cite{JD-McVean2005}. It
describes the history of each segment of the sampled sequences,
tracing back their ancestors in potentially distinct individuals. Such
segments, which have left descendants in the sample, are termed
\emph{ancestral}. Contained in the ARG is also the history of some
\emph{non-ancestral segments}, which did not leave a descendant in the
sample, but were once part of a sequence that contained both ancestral
and non-ancestral segments (Figure \ref{JD-fig1}D).

The characteristics of the ARG are determined by the demography of the
population (the history of population size changes), the recombination
landscape (where do the recombination events occur), but also the
selective forces acting on the sequence, as natural selection
influences the distribution of the number of descendants for each
individual, based on the nature of the sequences themselves.  While
the ARG contains the signature of the biological processes that shaped
the genome sequences, it is unfortunately not directly accessible. In
order to access the embedded information, it is necessary to model the
evolution of sequences in populations.

\subsection{The coalescent with recombination as a chronological
  process}

When modelling evolution, the most intuitive approach is to consider
the process chronologically, that is, to model the state of the system
generation after generation. One of the most simple models, the
so-called \emph{Wright--Fisher process},
\index{Wright--Fisher!process}\index{process!Wright--Fisher} 
considers that the gametes forming one generation are a random sample
of the gametes produced at the previous generation, that is,
reproduction is a purely random process where each individual has the
same a priori chance to contribute to the next generation. In
addition, the population has a finite, constant size. A similar model,
termed the \emph{Moran process}\index{Moran!process}, considers a
slightly different set-up with overlapping generations
\cite{JD-Moran1958}. The Wright--Fisher and Moran processes can both
accommodate recombination, modelled by randomly choosing a breakpoint
along the genome and exchanging the parental genetic segments (see the
contribution of Baake and Baake~\cite{JD-EBMB20} in this volume). In such processes,
the fate of a genetic variant is purely stochastic and governed only
by the population size.

When conditioning on a sample of the result of the evolutionary
process, a backwards-in-time modelling is used. Each sequence in the
sample represents a lineage, and the aim of the model is to determine
which lineages find a common ancestor in the past and when. Every time
two lineages \emph{coalesce} into a common ancestor, the number of
lineages to model is reduced by one, until the last common ancestor of
the sample is reached. This process of lineages merging backwards in
time is termed the \emph{coalescent}\index{coalescent}
\cite{JD-Kingman1982}.

The probability that two lineages merge at a given generation back in
time depends on the population size. When the population size is
constant in time, the number of generations until coalescence follows
a geometric distribution with parameter $\frac{1}{2\cdot
  N_e}$, where $N_e$ is the \emph{effective population size}.\index{effective!population size} 
Assuming a large $N_e$, this discrete-time coalescent is well
approximated by a continuous-time coalescent, where the divergence
time between two sequences follows an exponential distribution with
average $2\cdot N_e$ generations. For convenience, time is, therefore,
measured in ``coalescence'' units equal to $2\cdot N_e$, so that the
mean divergence time between two sequences in a sample is equal to 1.

In the \emph{coalescent with recombination}\index{coalescent!with recombination} 
process, recombination events are modelled in
addition to coalescence events
\cite{JD-Hudson1983}. Backwards-in-time, a lineage undergoing a
recombination event splits in two, the left and right sequences having
distinct ancestors. Since the rates of coalescence and recombination
events, at any time point, depend only on the current lineages, the
process is Markovian in time \cite{JD-Simonsen1997}. This property
enabled the development of simulation procedures and inference
methods, allowing the estimation of various parameters by integrating
over the unknown genealogy of a sample (e.g. \cite{JD-Adams2004,
  JD-Drummond2005}). Such methods, however, do not scale well with the
length of the modelled sequences, as the number of events in the
underlying ARG grows with the sequence length \cite{JD-Ethier1990},
preventing efficient integration even with Markov chains Monte-Carlo
\cite{JD-Stumpf2003}. These methods are, therefore, restricted to
small samples with relatively few loci.

\subsection{The coalescent with recombination as a sequential process}

Following the initial work by Simonsen and Churchill
\cite{JD-Simonsen1997}, Wiuf and Hein extended the two-loci model of
coalescence with recombination to multiple loci
\cite{JD-Wiuf1999}. The resulting process models the ARG sequentially
along the genome rather than chronologically. The resulting
\emph{sequential coalescent with recombination}\index{sequential!coalescent with recombination} aims at modelling the genealogy of
the sample at position $i$ given the genealogies at previous
positions. In fact, genealogies at two distinct positions in the
sequence are not independent: they are identical if no recombination
event occurred between the two positions since the last common
ancestor of the sample and can only differ if at least one
recombination event occurred. Despite this intuitive correlation
structure, the coalescent with recombination is not Markovian along
the sequence. Computing the probability distribution of the marginal
genealogy at a given position proved to be quite challenging because
of long-range dependencies, the genealogy at position $i$ depending
not only on the genealogy at position $i-1$, but on the genealogy at
all positions $1$ to $i-1$.

With the goal to simplify the likelihood calculation under the
coalescent with recombination, McVean and Cardin proposed an
approximation where certain types of coalescence events are ignored
\cite{JD-McVean2005}. An intuitive description of the simplified
process was provided by Marjoram and Wall \cite{JD-Marjoram2006}, who
recognised five types of recombination events on the ARG, based on the
type of segments in the parental sequences on both sides of the
recombination event (Figure~\ref{JD-fig1}D): type 1 events occur in
ancestral segments (events at $x_3$ and $x_4$ in Figure
\ref{JD-fig1}D) while types 2-5 occur in non-ancestral segments. Type
2 events occur in so-called \emph{trapped genetic
  material}\index{trapped genetic material} \cite{JD-Rasmussen2014},
that is, non-ancestral segments flanked on both sides by ancestral
segments (event at $x_1$ in Figure~\ref{JD-fig1}D). Events of types 3,
4, and 5 occur in non-ancestral segments only flanked by non-ancestral
segments on one or both sides (e.g. event at $x_2$ in Figure~\ref{JD-fig1}D).
Such events (3, 4 and 5) do not affect the sample
generated by the corresponding ARG, and therefore do not impact the
likelihood of the sample given the ARG. They can therefore be ignored
without introducing any additional hypotheses, see \cite{JD-Wiuf1999}
(types 4 and 5) and \cite{JD-Hudson2002} (types 3, 4 and 5). The
process of McVean and Cardin, which was further improved by Marjoram
and Wall \cite{JD-Marjoram2006} and Hobolth and Jensen
\cite{JD-Hobolth2014} also ignores type 2 recombination events, that
is, recombination events occurring in trapped genetic material
\cite{JD-Rasmussen2014}. Doing so implies ignoring potential
long-range dependencies between loci, and the distribution of samples
generated by this approximated process is, therefore, different from
that of the standard coalescent with recombination. The approximated
process, however, has the additional property that the distribution of
genealogies at position $i$ is only dependent on the genealogy at
position $i-1$, and is, therefore, Markovian along the sequence. Such
a process is referred to as the \emph{sequentially Markov coalescent
  (SMC)}\index{sequential!Markov coalescent} \cite{JD-McVean2005,
  JD-Marjoram2006}. Importantly, the SMC process generates samples
with patterns of genetic diversity that are very similar to the ones
generated by the full coalescent process \cite{JD-McVean2005}. The
SMC, in particular, can be seen as a first-order Markov approximation
of the true coalescent with recombination process
\cite{JD-Wilton2015}, and higher order extensions have been introduced
\cite{JD-Staab2015}. Furthermore, the Markov property enables very
efficient likelihood calculation using dynamic programming algorithms
to integrate over all ARGs. Such methodology comes from the field of
\emph{hidden Markov models (HMM)}\index{hidden Markov model}, which
we introduce in the next section.

\subsection{Coalescent hidden Markov models}

\begin{figure}[t]
    \begin{center}
       \includegraphics[width=\textwidth]{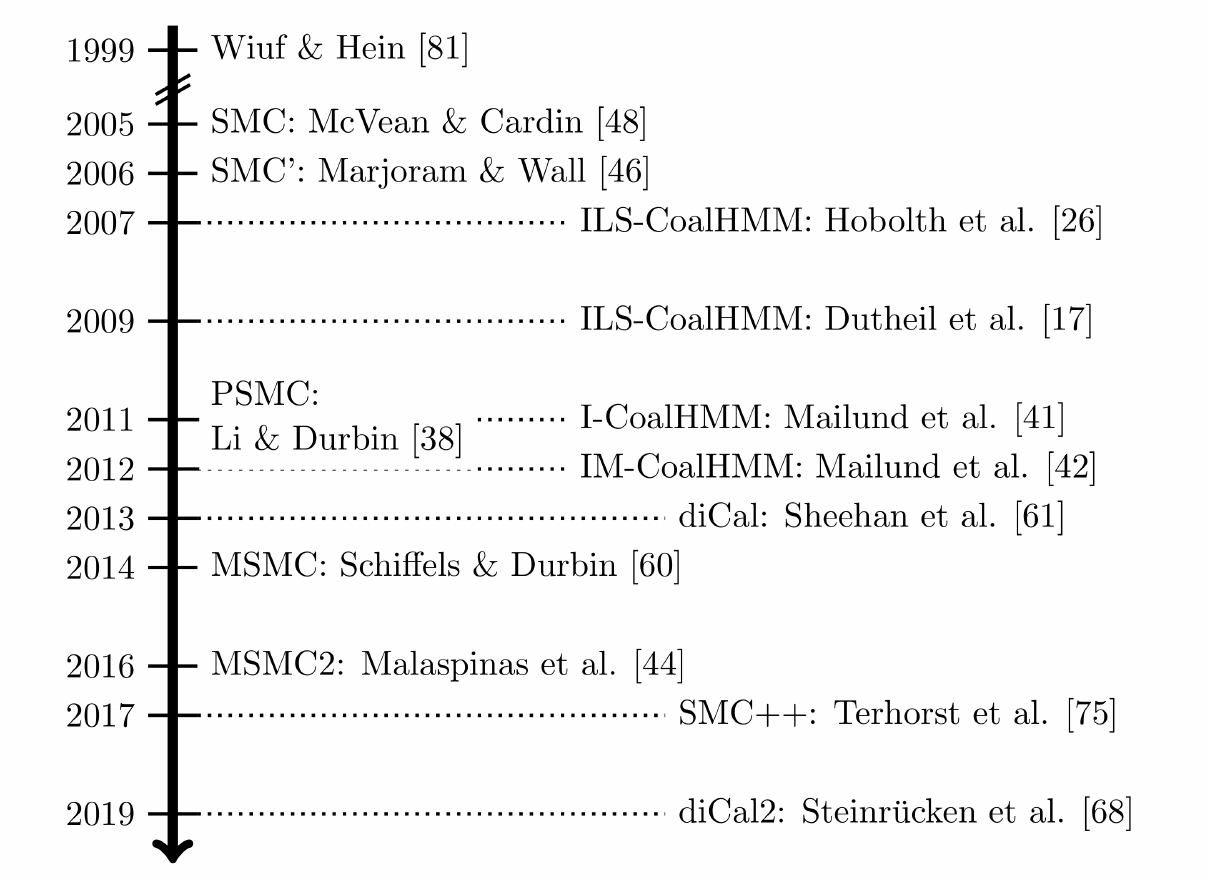}
    \end{center}
    \caption{\label{JD-fig2} Chronology of sequentially Markov
      coalescent (SMC) and coalescent hidden Markov models
      (CoalHMM). PSMC: pairwise SMC. MSMC: multiple SMC. ILS:
      incomplete lineage sorting. I: isolation model. IM: isolation
      with migration model.  }
\end{figure}

Because of the SMC approximation, likelihood calculation under a
coalescent with recombination process represents a classical
bioinformatic problem where the probability of an observed state in
the sequence depends on an unobserved state, which is then said to be
hidden. In the case studied here, the observed states are sequence
polymorphisms (between 2 or more individual sequences) and the hidden
states are the underlying marginal genealogies. HMMs have been broadly
used in sequence analysis \cite{JD-Durbin1998}. \emph{Coalescent
  hidden Markov models (CoalHMM)}\index{coalescent!hidden Markov model 
  (CoalHMM)} refer to HMMs where the hidden states are genealogies. It
was introduced by Hobolth et al \cite{JD-Hobolth2007} as a name of the
first model developed, which we introduce later in this section, but
was then extended to generally encompass a full class of models
\cite{JD-Spence2018} (Figure \ref{JD-fig2}).

We note as $\{\boldsymbol{\mathcal{A}}_i\}_{1\leq i \leq L}$ the
site-specific random variable of observed states in a sample of $M$
sequences of length $L$. Such states (noted
${\{A_g\}}_{1 \leq g \leq S}$) are, in the general case, a combination
of the four nucleotides $A$, $C$, $G$ and $T$ (one per modelled
sequence), with the possibility to additionally account for missing
data (coded as $N$), so that
${\{A_g\}}_{1\leq g \leq S} \in {\{A,C,G,T,N\}}^M$ and $S < 5^M$
because of symmetry relationships between trees making some of them
unidentifiable. 
We note as $\{x_i\}_{1 \leq i \leq L}$ a particular realisation of
$\{\boldsymbol{\mathcal{A}}_i\}_{1\leq i \leq L}$, that is, the
sequence data. Furthermore, we note as
$\{\boldsymbol{\mathcal{H}}_i\}_{1\leq i\leq L}$ the site-specific
random variable describing the marginal genealogies at each position
in the sequences. In the general case, such genealogies are rooted
trees with $M$ leaves.
 
In HMM terminology, the probabilities of observing the sequence data
$x_i$ at a given position $i$ given a realisation of
$\boldsymbol{\mathcal{H}}_i$,
$\operatorname{Pr}(\boldsymbol{\mathcal{A}}_i \mid
\boldsymbol{\mathcal{H}}_i )$, are called the \emph{emission
  probabilities}\index{emission probability}. 
$\boldsymbol{\mathcal{H}}_i$ is a random variable
that has a continuous distribution. To make likelihood calculations
tractable, this distribution is discretized, so that
$\boldsymbol{\mathcal{H}}_i$ can take a finite number $n$ of hidden
states, $\{H_j\}_{1\leq j \leq n}$. Under a discretised distribution
of hidden states, the emission probabilities for each position $i$,
hidden state $k$ can be more explicitly written as:
\begin{equation}
  e^{}_{i,k}(x) = \operatorname{Pr}(\boldsymbol{\mathcal{A}}^{}_i = 
x \mid \boldsymbol{\mathcal{H}}^{}_i = H^{}_k).
\end{equation}
We further introduce the so-called \emph{transition probability} of a
genealogy $H_j$ at position $i-1$ to a genealogy $H_k$ at position $i$
as
\begin{equation}
q^{}_{i,j,k} = \operatorname{Pr}(\boldsymbol{\mathcal{H}}^{}_i = 
H^{}_k \mid \boldsymbol{\mathcal{H}}^{}_{i-1} = H^{}_j).
\end{equation}
Given the set of emission and transition probabilities we can then
write the likelihood of the data by recursion. We define $F_{i,k}$ the
joint probability of the data (observed states) $x_1, \ldots, x_i$ at
positions 1 to $i$ and the ancestral genealogy $H_k$ at position $i$
as:
\begin{equation}\label{JD-eqn-forward}
F^{}_{i,k} = \operatorname{Pr}(x^{}_1, \ldots, x^{}_i, H^{}_k) = 
\left\{
\begin{array}{ll}
f_k & \mathrm{if}\, i = 0\\
e^{}_{i,k}(x^{}_i)\cdot \sum_j q^{}_{i, j, k}\cdot F^{}_{i-1,j}  & 
\mathrm{if}\, i > 0\\
\end{array}
\right.,
\end{equation}
where $\{f_k\}_{1 \leq k \leq n}$ denotes some initial
conditions. These conditions may be set to $1/n$, or to the stationary
distribution of the Markov chain (providing it exists). Equation
\eqref{JD-eqn-forward} is called the \emph{forward
  algorithm}\index{forward algorithm} and allows for the computation
of the likelihood of the sequences as
\begin{equation}
\mathcal{L} = \sum_k F^{}_{L,k}.
\end{equation}
This recursion is an example of dynamic programming, allowing for the
integration over all possible ARGs very efficiently, as it scales in
$\mathcal{O}(L\cdot n^2)$. The symmetry relationships in the
transition matrix and the relatively low frequency of the observed 
heterozygous states, however, allow for further improvement of the run
time of the forward algorithm
\cite{JD-Harris2014,JD-Terhorst2017,JD-Sand2013}.

The likelihood function depends on a set of parameters $\Theta$, which
includes the demographic model (effective population size and its
variation) and the recombination rate. More complex models can be
implemented, for instance allowing for population structure. These
parameters can affect either the emission probabilities $e_{i,k}(x)$,
the transition probabilities $q_{i,j,k}$, or both. The emission and
transition probabilities define the type of model used. Importantly,
most models assume that the process is homogeneous along the sequence,
so that both $e_{i,k}(x)$ and $q_{i,j,k}$ are actually independent of
$i$. In the following, we will review several examples of coalescent
HMM models.

\subsection{The two-genome case}

\begin{figure}[t]
    \begin{center}
        \includegraphics[width=\textwidth]{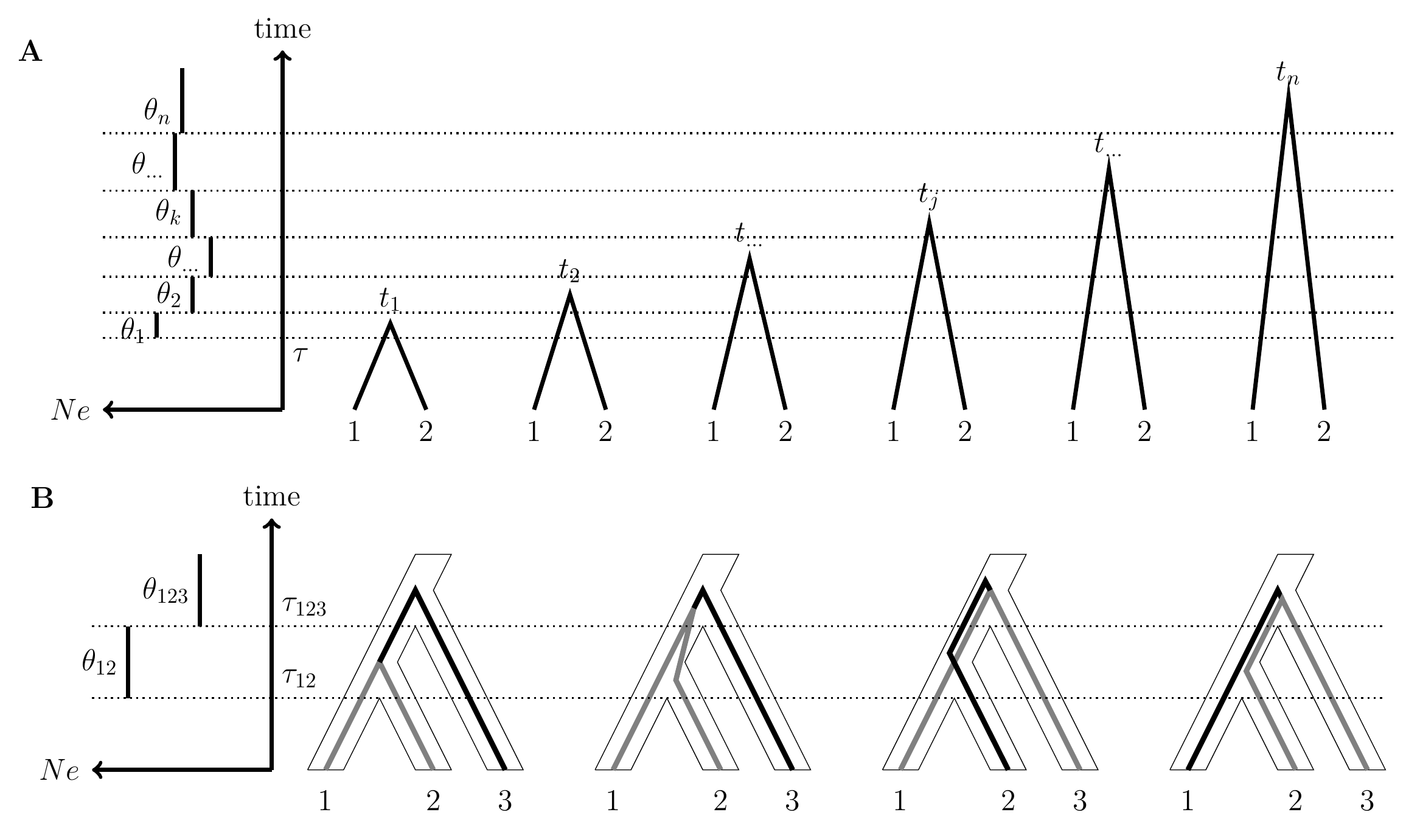}
    \end{center}
    \caption{\label{JD-fig3} Demographic models and hidden states for
      CoalHMM with two and three sequences. A: Pairwise SMC. Hidden
      states correspond to discretised divergence times between two
      sequences. Parameters of the model potentially contain a species
      divergence time $\tau$ and several epochs of constant effective
      population sizes $\theta_k$. The CoalHMM model of Mailund et
      al. \cite{JD-Mailund2011} uses only one epoch and $\theta$, the
      corresponding states $t_j$ being, therefore, drawn from the
      exponential distribution with mean $2\cdot\theta$, shifted by
      $\tau$. The PSMC model of Li and Durbin \cite{JD-Li2011},
      assumes a skyline model of multiple epochs, yet with individuals
      from the same population ($\tau = 0$). B: CoalHMM with three
      sequences and ILS. The hidden states correspond to four
      genealogies that differ both in time and order of the
      coalescence events. The model assumes constant but distinct
      ancestral effective population sizes $\theta_{12}$ and
      $\theta_{123}$, as well as the two species divergence times
      $\tau_{12}$ and $\tau_{123}$.  }
\end{figure}

When the genome sample consists only of two genomes, the marginal
genealogies have a more simple encoding consisting of a single
(continuous) number representing the time to the most recent common
ancestor (TMRCA) of the two sequences. The TMRCA can be further
discretised into $n$ hidden states, each represented by a mean value
$(t_j)_{1\leq j \leq n}$. The transition probabilities between the
hidden states can be calculated under the SMC. Variants of this model
were developed independently by Li and Durbin \cite{JD-Li2011} and
Mailund et al \cite{JD-Mailund2011}. In the latter, the two genomes
come from two distinct populations that diverged at a time $\tau$
units ago (Figure \ref{JD-fig3}A). The common ancestral population is
assumed to have a constant effective population size
$\theta_\text{anc}$. The TMRCA $(t_j)_{1\leq j \leq n}$ follows an
exponential distribution shifted by an amount of $\tau$. Mailund and
collaborators applied this model to the newly sequenced genomes of two
Orangutan subspecies in order to estimate their ancestral effective
population size and the time of their last genetic exchange. They
further extended this model to allow for a period of gene flow after
the initial separation of the two populations \cite{JD-Mailund2012}.

\begin{figure}[t]
    \begin{center}
        \includegraphics[width=\textwidth]{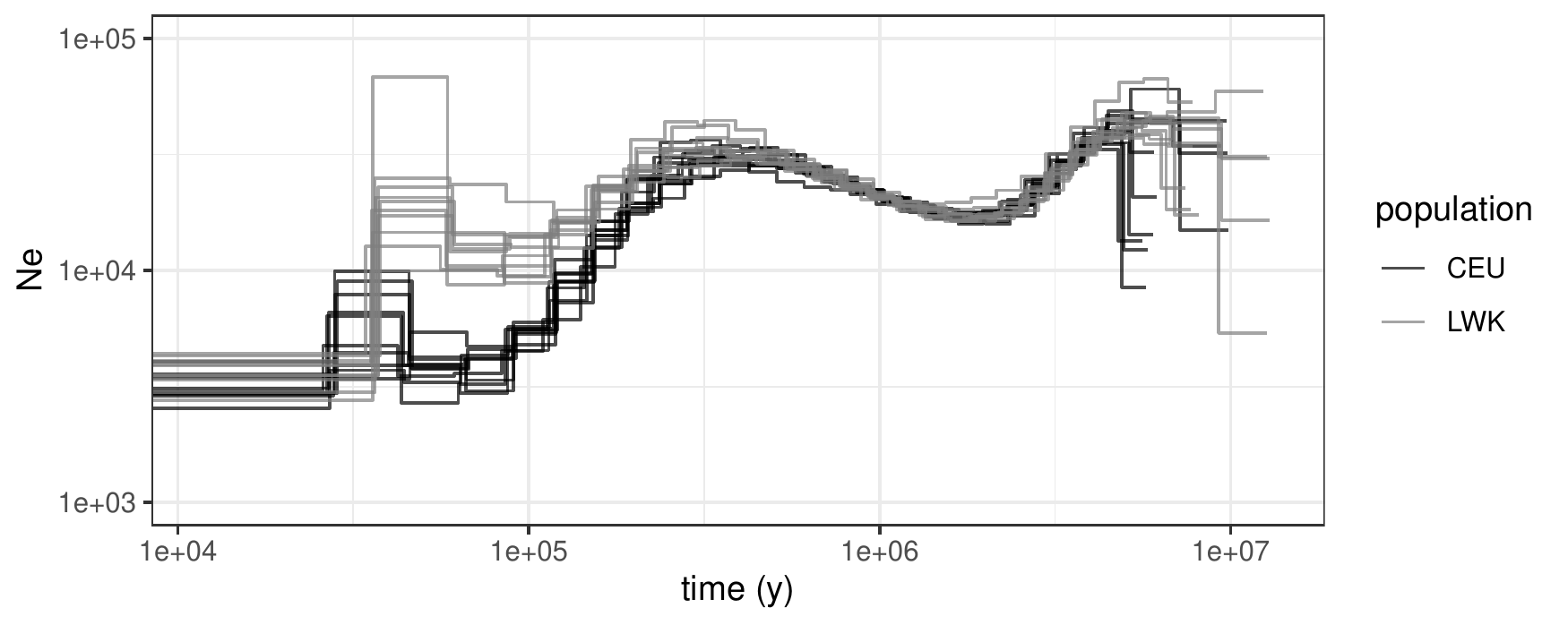}
    \end{center}
    \caption{\label{JD-fig4} Demographic inference under the PSMC. The
      MSMC2 software was used independently on 20 diploid individuals
      from the 1000 genomes project \cite{JD-1000G2010}, 10 from the
      CEU population (Utah residents with European ancestry) and 10
      from the LWK population (Luhya in Webuye, Kenya). MSMC2 was run
      on data from chromosome 9 only, with default parameters. The
      results show that individuals with European ancestry underwent a
      stronger bottleneck between 50 and 100 ky ago, corresponding to
      the out-of-Africa event.  }
\end{figure}

In the Li and Durbin model, named pairwise sequentially Markov
coalescent (PSMC), the two genomes come from a single population. This
approach was further improved by Schiffels and Durbin
\cite{JD-Schiffels2014}, who used a more accurate recombination
model. The authors consider a demographic model where the effective
population size is piecewise constant over a given number of epochs
(Figure \ref{JD-fig3}A). The parameters of the model consist of the
set of ancestral sizes, as well as the recombination rate, presumed to
be constant along the sequences. While the epochs of the demographic
model and the discretisation scheme used for the divergence time are
distinct aspects, it is convenient to have some overlap between the
two, providing that there are at least as many hidden states as epochs
(otherwise some parameters would become unidentifiable). Li and Durbin
proposed to consider one hidden state per epoch, so that each segment
is represented by one value of $t_{j}$ and one value of $\theta_j$
(Figure~\ref{JD-fig3}A).  By estimating one $\theta$ per epoch, the
PSMC model allows the reconstruction of a ``skyline'' plot where
population size varies in time, from present to the distant past
(Figure~\ref{JD-fig4}). This method was applied to data from the 1000
Genomes Project \cite{JD-1000G2010} in order to infer the demographic
history of distinct populations, which show the signature of the
out-of-Africa bottleneck.

The two approaches of Li and Durbin and Mailund et al. further differ
in their calculation of the emission probabilities. Focusing on the
population level and a relatively short time span of $\sim 1$ Myr, Li
and Durbin consider an \emph{infinite sites model}\index{infinite-sites!model}
where only one mutation per site can happen. They further
consider all types of mutations as equally probable and ignore the
biochemical nature of the underlying nucleotides. This reduces the
number of observed states to three types: homozygous (the two
sequences are identical at a given position), heterozygous (the two
sequences differ at a given position), and unknown (at least one
sequence has an unresolved state at that position). The emission
probabilities then take the simple form:
\begin{eqnarray}
e^{}_{i,j,\mbox{homozygous}} &=& \exp(-\theta\cdot t^{}_j)\\
e^{}_{i,j,\mbox{heterozygous}} &=& 1 - \exp(-\theta\cdot t^{}_j)\\
e^{}_{i,j,\mbox{unknown}} &=& 1,
\end{eqnarray}
where $\theta = 4\cdot N_e\cdot u$ denote the population mutation
rate, and $u$ the per nucleotide, per generation molecular mutation
rate.

Comparing genomes from two distinct (sub)species, therefore
potentially encompassing larger time scales, Mailund et al. used a
fully parametrised substitution model as used in parametric
phylogenetic reconstruction methods \cite{JD-Felsenstein2003}. The
mutation process is then a continuous time, discrete-state Markov
model with generator $U\!$, and the emission probabilities are given by
$\exp(U\cdot t_j)$ for each hidden state $j$.  In both models, the
emission probabilities only depend on the observed states and are
independent of the actual position in the sequence, assuming a
homogeneous mutation process along the genome.

\subsection{The three-genome case}

In 2007, Hobolth et al. introduced the first CoalHMM model
\cite{JD-Hobolth2007}, by modelling the possible genealogical
relationships between three species: ``$(A,B),C$'', ``$A,(B,C)$'' and
``$(A,C),B$'' (Figure~\ref{JD-fig3}B). Considering the two speciation
events that separate first the ancestor of $A$ and $B$ from the
ancestor of $C$, and then the ancestor of $A$ from the ancestor of
$B$, the probability that an individual sequence from $A$ coalesces
with a sequence from $B$ within the $AB$ ancestral species depends on
the ancestral population size and the time between the two speciation
events \cite{JD-Dutheil2012}. Backwards in time, if the two
corresponding lineages do not coalesce until the most ancient
speciation time, any of them can coalesce with a sequence from species
$C$ before coalescing with each other. This phenomenon, which results
in the marginal genealogy being incongruent with the phylogeny, is
termed \emph{incomplete lineage sorting (ILS)}\index{incomplete lineage sorting}. 
Using coalescent theory, Hobolth et al. derived
relationships between the transition probabilities and used them to
infer ancestral effective population sizes as well as the dates of
species divergence, the so-called speciation times. In this first
model the hidden states differ in tree topology and divergence
times. The first hidden state corresponds to the case where the two
lineages A and B coalesce within the ancestral population of A and B,
leading to a genealogy congruent with the phylogeny. The three other
topologies denote cases where the A and B lineages did not coalesce
within the AB ancestor, so that the three lineages A, B and C were
already present within the ancestral population of the three
species. These three states correspond to the cases where A and B, A
and C, or B and C coalesce first, respectively (Figure~\ref{JD-fig3}B).
The model assumes constant ancestral population sizes
and the divergence times for each topology are reduced to the averages
of the corresponding exponential distributions. The proportion of ILS
topologies directly depends on the time separating the speciation
events $\Delta_T$ and the effective size of the ancestral population
$\theta_\text{anc}$ \cite{JD-Dutheil2012}:
\begin{equation}
  \operatorname{Pr}(ILS) = \frac{2}{3}\exp
  \left(-2\cdot\frac{\theta^{}_\text{anc}}{\Delta^{}_T}\right),
\end{equation}
allowing the estimations of these parameters from the patterns of
topology variation.
The three-species CoalHMM model was applied to genome sequences of
Great Apes: Orangutan \cite{JD-Hobolth2011}, Gorilla
\cite{JD-Scally2012}, Bonobo \cite{JD-Prufer2012}, Baboons
\cite{JD-Rogers2019}, in order to infer the patterns of ILS and the
ancestral effective population sizes in this group of species
(reviewed in \cite{JD-Mailund2014}). It was also applied to species of
fungal pathogens where it was used to infer recombination rates
\cite{JD-Stukenbrock2011}.

\subsection{The multiple-genome cases}

With the development of sequencing technologies and the increasing
sample size of population genomic datasets, models able to extract the
genealogical information contained in multiple genomes are
needed. Building CoalHMM models with more than two or three sequences
poses, however, a computational challenge because of the underlying
combinatorics of marginal genealogies. The number of possible
topologies increases hyper-exponentially with the number $M$ of
sampled sequences, and there is an infinite number of possible
genealogies with a given topology due to the continuous nature of
branch lengths (divergence times). Further approximations are,
therefore, required to scale CoalHMM approaches with larger datasets.

\subsubsection{Using conditional sampling distributions}

In a series of articles
\cite{JD-Paul2011,JD-Steinrucken2013,JD-Sheehan2013,JD-Steinrucken2019},
Song and collaborators developed an approach based on the so-called
\emph{conditional sampling distribution} (CSD)
\index{sampling!distribution, conditional} introduced by Li and Stephens
\cite{JD-Li2003}. This approach stems from the chain rule of
conditional probabilities, allowing the expression of the likelihood
of a sample of $M$ sequences $S_1, \ldots, S_M$ as a product of
conditional likelihoods:
\begin{multline}
  \operatorname{Pr}(S^{}_1, S^{}_2, \ldots, S^{}_M \mid \Theta) =
  \operatorname{Pr}(S^{}_1 \mid S^{}_2, \ldots, S^{}_M,\Theta) \cdot
  \operatorname{Pr}(S^{}_2, \ldots, S^{}_M \mid \Theta)\\
  = \operatorname{Pr}(S^{}_1 \mid S^{}_2, \ldots, S^{}_M,\Theta) \cdot
  \operatorname{Pr}(S^{}_2 \mid S^{}_3, \ldots, S^{}_M \mid \Theta) \cdot
  \operatorname{Pr}(S^{}_3, \ldots, S^{}_M \mid \Theta)\\
  = \operatorname{Pr}(S^{}_M \mid \Theta) \prod_{k=1}^{M-1}
  \operatorname{Pr}(S^{}_k \mid S^{}_{k+1}, \ldots, S^{}_M,\Theta),
\end{multline}
where $\Theta$ denotes the parameter vector.  The conditional
likelihoods, however, are approximated, so that the resulting
likelihood is a product of approximate conditionals (PAC)
\cite{JD-Li2003}, which depends on the order by which the sequences
are treated in the product chain. This is usually accommodated by
permutations and averaging \cite{JD-Li2003}, or by a composite
likelihood approach such as the ``leave-one-out'' strategy
\cite{JD-Sheehan2013}:
\begin{equation}
  \operatorname{Pr}(S^{}_1, S^{}_2, \ldots, S^{}_M \mid \Theta)
  \simeq \prod_{i = 1}^M \operatorname{Pr}\big( S^{}_i \mid \left\{S^{}_j\right\}^{}_{j \neq i},\Theta\big).
\end{equation}
The CSD are computed under an SMC model, given a piecewise constant
demographic model, as in the PSMC. The model was further extended to
allow more complex demographic scenarios with population structure and
migration \cite{JD-Steinrucken2019}.

\subsubsection{Modelling the most recent coalescence events}

Schiffels and Durbin \cite{JD-Schiffels2014} developed the
\emph{multiple sequentially Markov coalescent (MSMC)}\index{sequential!Markov coalescent!multiple}, which models only the most recent
coalescent event in the sample. The underlying rationale was that the
PSMC is lacking resolution in the more recent past, due to the very
small number of mutations and coalescences happening in the most
recent epochs. Combining multiple samples, therefore, has the
potential to compensate for the lack of information in a single pair
of genomes. The MSMC approach is elegant as it reduces the
combinatorics of the hidden states to one continuous variable (which
is discretised, as in the PSMC): the time of coalescence, together
with the index of the two genomes in the sample that are coalescing,
bringing the number of hidden states to $n = {M\choose 2} \cdot K$,
where $K$ is the number of discrete classes used for the distribution
of divergence times. The efficiency of the MSMC approach is, however,
paradoxical: by gaining resolution in the present as the sample size
increases, the method progressively loses power as the number of
modelled sequences becomes larger (see \cite{JD-Dutheil2017} for an
illustration). In practice, the authors showed that for a human
dataset, the maximum resolution is obtained for eight haploid genomes
\cite{JD-Schiffels2014}.

\subsubsection{Using a composite likelihood}\label{JD-subsubsection-composite}

In the MSMC2 approach \cite{JD-Malaspinas2016, JD-Wang2019}, Schiffels
and collaborators proposed to approximate the likelihood of a sample
of $M$ genomes by independently considering all pairs of genomes. The
likelihood of the sample is then approximated by the product of all
pairwise likelihoods, each computed under the PSMC model. While the
pairwise likelihoods are exact under the SMC, the likelihood of the
sample is an example of \emph{composite likelihood}\index{likelihood!composite} \cite{JD-Larribe2011}:
\begin{equation}
  \operatorname{Pr}(S^{}_1, S^{}_2, \ldots, S^{}_M \mid \Theta) \simeq
  \prod_{i = 1}^{M - 1} \prod_{j = i+1}^M \operatorname{Pr}
  (S^{}_i, S^{}_j \mid \Theta). 
\end{equation}
The likelihood here is an approximation since the divergence times
between pairs of sequences in a genealogy are not independent. The
MSMC2 approach is therefore better described as a ``multiple pairwise
SMC''. It was shown to display good resolution in both the past and
present time, efficiently making use of the increasing quantity of
signal as the sample size increases.

\subsubsection{Augmenting the PSMC with site frequency spectra}\index{site-frequency spectrum}

While approaches like diCal and MSMC2 allow for the efficient
modelling of the evolution of multiple genomes, they are intrinsically
limited as the computational cost become prohibitive for samples of
more than one or two dozen genomes (at least for genomes with a size
of the order of that of humans). Terhorst and colleagues introduced a
hybrid approach combining the power of the SMC, which makes efficient
use of linkage patterns, with that of classical site frequency
spectrum (SFS) based approaches \cite{JD-Terhorst2017}. This modelling
framework, termed SMC++, considers a ``focus'' diploid individual that
is modelled with a PSMC approach. The observed states are then
augmented by taking into account additional genomes to compute an
SFS. The emission probabilities are calculated as the probability of
observing the local SFS given the genealogy at the focus individual,
and the authors proposed an approach to compute such a conditional
site frequency spectrum (CSFS). The resulting SMC++ model can
accommodate hundreds of individual genomes. Another innovation
introduced in this approach is the abandonment of the ``skyline''
model of piecewise constant effective population size in favour of a
spline model. While divergence times are still discretised, the
corresponding times for each category are derived from a spline curve
whose parameters are estimated. This reduces the number of parameters
to estimate and ensures smoother inferred demographies. The SMC++
approach has been applied to human data as well as other species,
including Drosophila and Zebra finch \cite{JD-Terhorst2017}.

\section{Heterogeneity of processes along the genome}
\label{JD-section-het}

In all models that we evoked so far, evolutionary processes have been
considered to be homogeneous along sequences. In this section, I
review evidence that these assumptions are at odds with our current
knowledge of the biology of genomes.

\subsection{Variation of the recombination rate}

Recombination rates can vary extensively between species
\cite{JD-Stapley2017}, between sexes \cite{JD-Lenormand2005} and
within genomes. At the molecular scale, multiple levels can be
distinguished: recombination rate correlates negatively with
chromosome size, a pattern attributed to the mechanism of meiosis and
crossing-over interference \cite{JD-Kaback1996}. Given that the rate
of crossing-over events is low, this leads to a higher recombination
rate in small chromosomes. Recombination rates vary also within
chromosomes: in Primates, it is generally higher at the start and end
of the chromosomes (the so-called telomeric regions)
\cite{JD-Stevison2016}, while in Drosophila the opposite pattern is
observed \cite{JD-Comeron2012,JD-Chan2012}. In many species
recombination events have an increased chance to occur in particular
regions, called hotspots
\cite{JD-Petes2001,JD-Myers2005,JD-Stukenbrock2018} (but see
\cite{JD-Wallberg2015} for a counterexample).

The variation of recombination rate has two types of consequences on
the patterns of sequence diversity. Because the molecular mechanisms
of recombination are tightly linked to DNA repair, recombination
itself can be mutagenic and locally increase sequence variability
\cite{JD-Lercher2002,JD-Alves2017}. Furthermore, in many species, the
repair mechanisms involve gene conversion between homologous
sequences, as one chromosome is used as a template to repair the other
one. However, this mechanism is biased in many species: in the case of
heterozygous positions, the ``C'' or ``G'' nucleotides are preferred
over ``A'' and ``T'' nucleotides, potentially resulting in large scale
variations of GC content \cite{JD-Duret2009} mirroring the variations
in recombination rate. The recombination rate also has indirect
effects on genetic diversity: because it breaks down genetic linkage,
recombination counteracts the reduction of diversity at sites linked
to loci under selection, both negative (background selection
\cite{JD-Charlesworth1993}) and positive (genetic hitch-hiking
\cite{JD-Barton2000}). By modulating the local effective population
size, variation of recombination rate along the genome has a strong
impact on the underlying genealogy.

\subsection{Variation of the mutation rate}

Finally, the rate at which mutations occur can vary extensively along
the genome \cite{JD-Baer2007}. Mutations can occur via direct
modification of the DNA or indirectly, via errors in the replication
or repair mechanisms. Particular sequence motifs, such as CpG
dinucleotides are known to undergo comparatively higher mutation
rates, via the methylation of the cytosine, which is then deaminated
into a thymine, leading to a TpG dinucleotide. In addition to the
potentially mutagenic effect of recombination, which also plays a role
in the repair of DNA damage, the replication machinery itself is
error-prone. This error rate is position dependent: it increases with
the replication time, being lower close to the origins of replication
\cite{JD-Stamatoyannopoulos2009, JD-Agier2012, JD-Weber2012}. Under a
neutral scenario, the mutation process is independent of the
coalescent process \cite{JD-Hudson1983}, and, therefore, has no impact
on the underlying genealogies. Yet, for inference models, mutation
rate variation acts as a confounding factor, as a high sequence
divergence can be either explained by an ancient coalescence time or a
high mutation rate. In CoalHMM models, the mutation rate will have an
impact on the emission probabilities, that is, the probability of
observing the observed sequence diversity given a genealogy.

\section[]{Existing approaches to account for spatial heterogeneity}

In this section, I review the approaches that have been developed to
cope with the heterogeneity of evolutionary processes acting along the
genome.

\subsection{Inferring sequential heterogeneity alone}

A large body of work is built on the idea that, if a parameter affects
certain patterns of genetic diversity, it should be possible to use
these patterns to recover the underlying variation of the
parameter. The most studied case in this respect is the recombination
rate, via its impact on linkage disequilibrium. Given an a priori
known demographic scenario, it is possible to compute the likelihood
of the data for any given recombination rate, and use it to estimate
the most likely recombination rate value. Due to the complexity of the
underlying model, however, approximations are required to apply these
methods to large genomic datasets. McVean, Awadalla and Fearnhead
\cite{JD-McVean2002} introduced the use of a composite likelihood,
approximating the full likelihood by the product of the likelihoods of
all pairs of positions within a minimum distance of each other. This
approach is the basis of several popular methods for recombination
rate inference such as LDhat \cite{JD-Auton2012} and LDhelmet
\cite{JD-Chan2012}. Further developments of these models allowed for
the incorporation of variable population sizes \cite{JD-Kamm2016,
  JD-Spence2019}. The underlying demography, however, has to be
estimated independently from the data.  Li and Stephens
\cite{JD-Li2003}, on the other hand, used the conditional sampling
distribution and the product of approximate conditionals (PAC) to
approximate the likelihood. An application of this method also
includes the reconstruction of haplotypes from genotypic data, a
problem known as \emph{phasing}\index{phasing} \cite{JD-Stephens2005}.

\subsection{Inference using piecewise-homogeneous processes}

The most simple approach to infer heterogeneous processes along the
genome while jointly accounting for demography is to use a
window-based approach, consisting of dividing the genome into segments
of fixed sizes and estimating model parameters independently in each
resulting window. This strategy was used by Stukenbrock et
al. \cite{JD-Stukenbrock2011} to use the patterns of ILS and a CoalHMM
model to estimate the recombination rate in 100 kb windows along the
genome of the fungal pathogen \textit{Zymoseptoria tritici}. In most
cases, however, SMC models require long genome sequences to be able to
confidently estimate parameters, and cannot be run in windows of small
sizes, in particular for models at the population level. Furthermore,
window-based approaches raise the issue of the window size and
boundaries to use.

\subsection{Using sequentially heterogeneous simulation procedures}

While the computation of the likelihood of the data under a
sequentially heterogenous process is notoriously difficult, simulating
under the corresponding model can be comparatively easy. Software like
the Markovian coalescent simulator (MaCS) \cite{JD-Chen2009}, the
sequential coalescent with recombination model (SCRM)
\cite{JD-Staab2015}, fastsimcoal \cite{JD-Excoffier2011} and MSprime
\cite{JD-Kelleher2016} allow the simulation of population genomic data
sets under models with variable recombination rate. Owing to their
high computational efficiency, they can be used within an approximate
Bayesian computation (ABC) framework in order to estimate demographic
parameters under realistic recombination models
\cite{JD-Wegmann2010}. This possibility, however, has to date been
underexploited, as demographic inference is so far conducted with data
simulated under a homogeneous recombination landscape (see for
instance \cite{JD-Li2012}). While no ABC method has been developed
with the goal to infer the variation of population genomic parameters
along the genome, Gao et al.~\cite{JD-Gao2016} introduced a machine
learning approach to infer recombination rates. The underlying
simulations, however, are conducted under a model with constant
recombination rate.
 
\subsection{A posteriori inference of heterogeneous processes}

The HMM methodology allows, via the forward algorithm, to compute the
likelihood of the data given a specified demographic model by
efficiently integrating over the unknown underlying ARG. The HMM
toolbox further allows for the computation of the a posteriori
probability of each marginal genealogy for each position
\cite{JD-Dutheil2017}:
\begin{equation}
  \operatorname{Pr}(\boldsymbol{\mathcal{H}}^{}_i = H^{}_j
  \mid x^{}_1, \ldots, x^{}_L) =
  \frac{\operatorname{Pr}(x^{}_1, \ldots, x^{}_L,
    \boldsymbol{\mathcal{H}}^{}_i = H^{}_j)}{\operatorname{Pr}
    (x^{}_1, \ldots, x^{}_L)}. 
\end{equation}
The denominator of the ratio is the likelihood of the data,
$\mathcal{L}$. In order to compute the numerator, we need to introduce
the so-called \emph{backward algorithm}\index{backward!algorithm}
\cite{JD-Durbin1998}:
\begin{equation}\label{JD-eqn-backward}
  B^{}_{i,j} = \operatorname{Pr}(x^{}_{i+1}, \ldots, x^{}_L
  \mid \boldsymbol{\mathcal{H}}^{}_i = H^{}_j) = \left\{
\begin{array}{ll}
1 & \mathrm{if}\, i = L\\
  \sum_k e^{}_{i+1,k}(x^{}_{i+1})\cdot q^{}_{i+1, j, k}\cdot
  B^{}_{i+1,k} & \mathrm{if}\, i < L , \\
\end{array}
\right.
\end{equation}
The posterior probability of hidden state $H_j$ can therefore be
computed as
\begin{equation}
  \operatorname{Pr}(\boldsymbol{\mathcal{H}}^{}_i = H^{}_j \mid x^{}_1,
  \ldots, x^{}_L) =
\frac{F^{}_{i,j}\cdot B^{}_{i,j}}{\mathcal{L}}. 
\end{equation}
This formula allows for the reconstruction of the most probable
marginal genealogy at each position $i$ by taking the maximum
posterior probability
\begin{equation}
  \widehat{\boldsymbol{\mathcal{H}}}^{}_i = \underset{j}{\operatorname{arg\,max}}\left(\operatorname{Pr}(\boldsymbol{\mathcal{H}}^{}_i = H^{}_j \mid x^{}_1, \ldots, x^{}_L)\right), 
\end{equation}
a procedure called \emph{posterior decoding}\index{posterior decoding}.
The posterior decoding is performed after fitting the
model parameters by maximizing the likelihood. It is therefore an
example of \emph{empirical Bayesian}\index{empirical Bayesian}
inference \cite{JD-Maritz2018}.

Posterior probabilities of marginal genealogies can also be used to
obtain posterior estimates of biological quantities of interest,
accounting for the uncertainty on the underlying genealogy. The
posterior mean estimate $\widehat{\lambda}_i$ at position $i$ of a
property $\Lambda(H_j)$ can be obtained by
\begin{equation}\label{JD-eqn-postav}
  \widehat{\lambda}^{}_i = \sum_j \operatorname{Pr}
  (\boldsymbol{\mathcal{H}}^{}_i = H^{}_j \mid x^{}_i, \ldots, x^{}_L)
  \cdot\Lambda(H^{}_j).
\end{equation}
If $\Lambda$ is the coalescence time between two sequences, this
formula can be used to get posterior estimates of sequence divergence
along the genome \cite{JD-Palamara2018}. More complex examples of
functions include, for instance, whether $\boldsymbol{\mathcal{H}}_i$
is distinct from $\boldsymbol{\mathcal{H}}_{i-1}$, or, in other words,
whether a recombination event occurred between positions $i$ and
$i-1$. This allows for the reconstruction of a recombination map,
integrating over the ARG. Such approach was notably used by Munch et
al \cite{JD-Munch2014} to reconstruct the recombination map of the
human-chimpanzee ancestor. Posterior estimates are rather robust to
the specified input model and can therefore offer a powerful approach
to infer aspects of the process that are not directly accounted for by
the model. However, because some model properties are intrinsically
confounded, such as local divergence and mutation rate, ignoring
spatial heterogeneity might result in biased inference
\cite{JD-Barroso2019}.

\section{The integrative sequentially Markov coalescent}

In order to account for heterogeneous processes along the genome, we
recently developed the integrative sequentially Markov coalescent
(iSMC) \cite{JD-Barroso2019}, an extension of the SMC. In this
framework, parameters of the original SMC vary along the genome in a
Markovian manner, allowing for the modelling of genome heterogeneity
in addition to demographic processes. I illustrate this approach with
results from a recent application of this framework to infer
recombination landscapes and further discuss possible extensions.

\subsection{The Markov-modulated sequentially Markov coalescent}

Whilst the framework can be applied to cases where more than one
parameter varies along the genome, for simplicity, we here consider
the case where one parameter only varies, which we label
$\boldsymbol{\mathcal{R}}$, $\boldsymbol{\mathcal{R}}_i$ denoting the
values of $\boldsymbol{\mathcal{R}}$ at position $i$ in the
sequences. We assume that $\boldsymbol{\mathcal{R}}$ follows an a
priori known discrete distribution with $n^R$ categories, each with
mean value $R_k$, with $1 \leq k \leq n^R$. In the iSMC framework, the
transition and/or emission probabilities are functions of
$\boldsymbol{\mathcal{R}}$ and are, therefore, noted as
$e^\text{SMC}_{i,k}(x, R)$ and $q^\text{SMC}_{i,j,k}(R)$,
respectively. The key assumption is then to consider that the
variation of $\boldsymbol{\mathcal{R}}$ along the genome can be
modelled as a Markov model, that is, there is a matrix of
probabilities $q^R$ defined as
\begin{equation}
  q^R_{i, j,k} = \operatorname{Pr}(\boldsymbol{\mathcal{R}}^{}_i = R^{}_k
  \mid \boldsymbol{\mathcal{R}}^{}_{i-1} = R^{}_j).
\end{equation}
The forward recursion of the CoalHMM can then be written as
\begin{multline}
  F^\mathrm{SMC}_{i,j,k} = \operatorname{Pr}(x^{}_1, \ldots, x^{}_i, H^{}_j, R^{}_k) = \\
  \left\{
\begin{array}{ll}
f^{}_j \cdot f^R_k & \mathrm{if}\, i = 0\\[2mm]
  e^{\mathrm{SMC}}_{i,j}(x^{}_i, R^{}_k)\cdot\sum_u \sum_v q^{\mathrm{SMC}}_{i, u, j}
  (R^{}_k)  \cdot q^R_{i, v, k}\cdot F^{\mathrm{SMC}}_{i-1,u,v}  & \mathrm{if}\, i > 0 ,\\
\end{array}
\right.
\end{multline}
where $f^R_k$ denotes the a priori probability
$\operatorname{Pr}(\boldsymbol{\mathcal{R}} = R_k)$.  Because the SMC
now depends on a parameter that itself follows a Markov process, the
resulting process can be described as a Markov-modulated Markov chain
(MMMC). As an MMMC is itself a Markov process \cite{JD-Galtier2004},
we can rewrite the forward recursion as:
\begin{equation}\label{JD-eqn-forward-ismc}
  F^{\text{iSMC}}_{i,k} = \operatorname{Pr}(x^{}_1, \ldots, x^{}_i,
  H^{\text{iSMC}}_k) = \left\{
\begin{array}{ll}
f^{\text{iSMC}}_k & \mathrm{if}\, i = 0\\[2mm]
  e^{\text{iSMC}}_{i,k}(x^{}_i)\cdot\sum_j q^{\text{iSMC}}_{i, j, k}\cdot
  F^{\text{iSMC}}_{i-1,j}  & \mathrm{if}\, i > 0 .\\
\end{array}
\right. 
\end{equation}
In the iSMC hidden Markov model, the hidden states $H^\text{iSMC}$
consist of all possible pairs of genealogies and
$\boldsymbol{\mathcal{R}}$: $(R_a, H_b)$, with $1 \leq a \leq n^R$ and
$1 \leq b \leq n$. The emission probabilities $e^\text{iSMC}_{i,j}(x)$
now depend on the pair $(R,H)_j$, and the initial probabilities of the
hidden states are $f^\text{iSMC} = f^R \otimes f$. Similarly, the
transition probabilities of the Markov-modulated SMC (MMSMC) can be
written as a function of the transition probabilities of the two
Markov chains:
\begin{equation}
q_i^{\text{iSMC}} = \left(
\begin{array}{ccc}
  q^R_{i,1,1} \cdot q_i^{\text{SMC}}(R^{}_1) & \cdots & q^R_{i,1,n^R}\cdot
     q_i^{\text{SMC}}(R^{}_1) \\
\vdots & \ddots & \vdots \\
  q^R_{i,n^R,1} \cdot q_i^{\text{SMC}}(R^{}_{(n^R)}) & \cdots & q^R_{i,n^R,n^R}
                      \cdot q_i^{\text{SMC}}(R^{}_{(n^R)})
\end{array}\right) .
\end{equation}
(As we consider the process modelling the variation of
$\boldsymbol{\mathcal{R}}$ to be itself homogeneous along the genome,
we have
$\forall i_1, i_2,\;q^{\text{iSMC}}_{i_1, j,k} = q^{\text{iSMC}}_{i_2,
  j,k} = q^{\text{iSMC}}_{j,k}$.)  The iSMC model can therefore be
analysed with standard HMM methodology, just like the homogeneous
SMC. The number of hidden states, however, is now $n^R\cdot n$,
meaning that the complexity of the likelihood calculation becomes
$\mathcal{O}\big(L\cdot (n \cdot n^R)^2\big)$.

The iSMC model adds relatively few extra parameters to the SMC: the
transition probabilities $q^R$, which can be reduced to one parameter
(see below), and parameters of the a priori distribution of
$\boldsymbol{\mathcal{R}}$.  While parameters of the distribution of
$\boldsymbol{\mathcal{R}}$ are generally not of direct biological
interest, the posterior decoding of the HMM allows for the inference
of the underlying landscape of the heterogeneous parameter. Distinct
decoding procedures can be performed in the case of Markov-modulated
HMMs:
\begin{enumerate}
\item A full decoding, where the most likely pair $(R,H)$ at each
  position is reconstructed:
    \begin{equation}
      \widehat{\left(\boldsymbol{\mathcal{R}}, \boldsymbol{\mathcal{H}}\right)}^{}_i = \underset{j,k}{\operatorname{arg\,max}}\left(\operatorname{Pr}
        (\boldsymbol{\mathcal{H}}^{}_i = H^{}_j, \boldsymbol{\mathcal{R}}^{}_i
        = R^{}_k \mid x^{}_i, \ldots, x^{}_L)\right),     
    \end{equation}    
  \item A partial decoding of genealogies, where the most likely
    genealogy is inferred, summing over all heterogeneous parameters:
    \begin{equation}
    \hat{\boldsymbol{\mathcal{H}}^{}_i} = \underset{j}{\operatorname{arg\,max}}\Big(\sum_k\operatorname{Pr}(\boldsymbol{\mathcal{H}}^{}_i = H^{}_j, \boldsymbol{\mathcal{R}}^{}_i = R^{}_k \mid x^{}_i, \ldots, x^{}_L)\Big),     
    \end{equation}
  \item A posterior mean estimation of the heterogeneous
    variable. Setting $\Lambda(H_j, R_j) = R_j$ and applying equation
    \ref{JD-eqn-postav}, we get:
    \begin{multline}\label{JD-eqn-post3}
      \hat{\boldsymbol{\mathcal{R}}^{}_i} = \sum_j \sum_k \operatorname{Pr}
      (\boldsymbol{\mathcal{H}}^{}_i = H^{}_j, \boldsymbol{\mathcal{R}}^{}_i
      = R^{}_k \mid x^{}_i,  \ldots, x^{}_L)\cdot R^{}_k\\
     = \sum_k R^{}_k \cdot\sum_j \operatorname{Pr}(\boldsymbol{\mathcal{H}}^{}_i = H^{}_j, \boldsymbol{\mathcal{R}}^{}_i = R^{}_k \mid x^{}_i,  \ldots, x^{}_L).\\
    \end{multline}
\end{enumerate}
The partial decoding of genealogies enables the reconstruction of the
ARG while accounting for the heterogeneity of the SMC along the
genome. The posterior estimates of the heterogeneous variable allows
the reconstruction of the variation along the genome while accounting
for the genealogy and its uncertainty. In the next section, we apply
this framework to model the variation of the recombination rate along
the genome.

\subsection{A case study: inference of recombination rate variation}

Recombination is the best documented heterogeneous process along the
genome. It can be measured experimentally or indirectly using genomic
approaches (reviewed in \cite{JD-Auton2012}). In the context of the
sequential coalescent and the SMC approximation, the local
recombination rate affects the probability of transition from one
genealogy to another, which increases with higher recombination
rates. To model variable recombination rates in iSMC, we considered
that the local population recombination rate
$\rho = 4\cdot N_e\cdot r$, where $r$ is the molecular recombination
rate in cM / bp per generation, is the product of a genome average
$\rho_G$ and a local modifier $r^\rho$. This modifier follows a prior
discrete distribution of mean 1 and with $n^\rho$ categories, for
instance a discretised Gamma distribution with shape parameter
$\alpha$. We further considered a simple model for the transition
probabilities between the $r^\rho$ classes, assuming equal
probabilities of change, $\gamma$. The transition probability matrix
$q^\rho$ takes the form:
\begin{equation}
q^\rho = \left(\begin{array}{cccc}
1 - \gamma & \gamma / (n^\rho - 1) & \cdots & \gamma / (n^\rho - 1)\\
\gamma / (n^\rho - 1) & 1 - \gamma & \ddots  & \vdots \\
\vdots & \ddots & \ddots & \gamma / (n^\rho - 1) \\
\gamma / (n^\rho - 1) & \cdots & \gamma / (n^\rho - 1) & 1 - \gamma\\
\end{array}\right).
\end{equation} 
Using the forward equation~\ref{JD-eqn-forward-ismc}, we can compute
the likelihood of the parameters of the iSMC model. Using optimisation
procedures, estimates of the $\alpha$ and $\gamma$ parameters together
with the average $\rho_G$ and the demography parameters can be
obtained by maximizing the likelihood function. The likelihood
calculation can also be used to perform model comparisons and test for
the heterogeneity of the coalescent process along the genome, for
instance using Akaike's information criterion
(AIC).

\begin{figure}[t]
    \begin{center}
        \includegraphics[width=\textwidth]{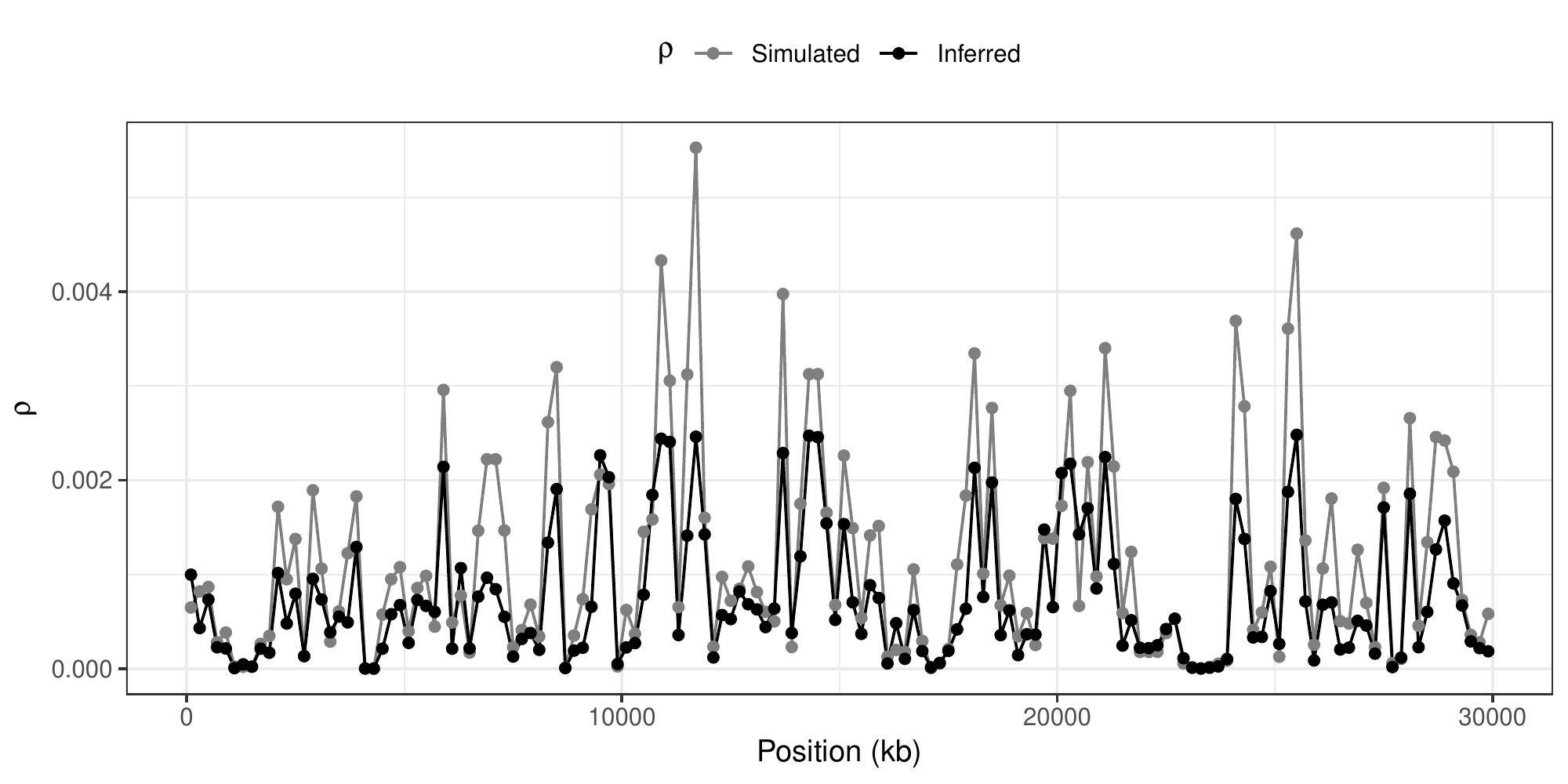}
    \end{center}
    \caption{\label{JD-fig5} Posterior estimates of recombination
      rates from a single diploid genome using iSMC. A 30 Mb region
      was simulated using the SCRM program \cite{JD-Staab2015} and a
      variable recombination rate, and then inferred with iSMC, as
      described in \cite{JD-Barroso2019}. Recombination rates were
      averaged in windows of 200 kb.  }
\end{figure}

A posterior decoding approach (equation \ref{JD-eqn-post3}) can then
be used to obtain estimates of site-specific recombination
rates. Simulations under controlled recombination landscape and
demography can be used to assess the accuracy of the iSMC inference
\cite{JD-Barroso2019}. Figure \ref{JD-fig5} shows that iSMC recovers
the underlying recombination landscape with good accuracy, despite
generally underestimating high recombination rates. A possible
explanation for this is the discretisation procedure, as the posterior
mean estimate is bounded by the class with the highest mean
recombination value. Allowing for more recombination classes allows
for a wider range of values and can potentially reduce this bias, at
the cost of increasing the running time and memory usage. Because it
can recover the recombination landscape from a single diploid only,
the iSMC model was used to generate recombination maps from extinct
hominids from their ancient genome sequences \cite{JD-Barroso2019}.

\subsection{Extension of iSMC: multiple genomes and multiple
  heterogeneous parameters}

The Markov-modulated Markov model framework underlying the iSMC
approach can be applied to other SMC models, such as the MSMC
\cite{JD-Schiffels2014}. Hidden states of the resulting CoalHMM are
combinations of TMRCA, pairs of genome indices undergoing the most
recent coalescent event, and classes of heterogenous parameters such
as the recombination
rate. 
Extension to multiple genomes can also be achieved using a composite
likelihood approach, as implementated in MSMC2
(see~\ref{JD-subsubsection-composite}). The likelihood of the dataset
is then approximated by the product of the likelihoods of each pair of
genomes, which are modelled separately with their own process. In the
case of the iSMC approach, this implies considering that the
heterogeneous parameters vary independently along each pair of
genomes; for a model with variable recombination rate, this is
equivalent to estimating a distinct recombination map for each pair of
genomes. This assumption is clearly incorrect for the vast majority of
positions within genomes from the same population. Extensions
enforcing a common map while allowing the coalescent processes to be
independent will, therefore, be instrumental in efficiently scaling
the iSMC approach to larger sample sizes.

The iSMC framework further allows the joint modelling of the variation
of multiple parameters along the genome, such as the mutation and
recombination rates. In such approach, the hidden states are a
combination of TMRCA and classes for each discretised parameter. The
addition of any additional heterogeneous parameter multiplies the
complexity of the likelihood calculation by ${n^\phi}^2$, where
$n^\phi$ is the number of discrete classes considered for the
parameter distribution. Besides the increased complexity,
identifiability issues may also arise, since local patterns of
diversity may be equally explained by variation of demography,
recombination rate or mutation rate.

However, when these parameters vary at a scale larger than the
variation of the TMRCA along the genome and when very large genome
sequences are analyzed, increasingly complex models may be
successfully fitted.

\section{Conclusions}

The availability of complete genome data opened the floodgates for the
detailed inference of the demographic history of species. A new
generation of coalescent-based models permits the extraction of
demographic signal from the patterns of genetic linkage along
sequences. Such models, however, largely ignore fundamental aspects of
genome biology, that is, that processes such as recombination and
mutation are highly heterogeneous along genomes. Extending these
approaches to account for such heterogeneity not only potentially
improves demographic inference, but also allows to reconstruct the
underlying genomic landscape.

\subsection*{Acknowledgement}

I would like to thank Gustavo V.~Barroso for critical reading of this
manuscript and for providing the simulation data plotted in
Figure~\ref{JD-fig5}. I am deeply indebted to Ellen Baake, as well as
Jeffrey~P.~Spence and an anonymous reviewer for their careful reading
of the manuscript, for finding several mistakes and typos, and for
their suggestions on how to improve its clarity.

\end{document}